\documentclass[a4paper, amsfonts, amssymb, amsmath, reprint, showkeys, footinbib, twoside]{revtex4-1}
\usepackage[english]{babel}
\usepackage[utf8]{inputenc}
\usepackage[margin=.5in]{geometry}
\usepackage{enumitem}
\usepackage[pdftex,pdftitle={PRL Yukawa Couplings},pdfauthor={Berglund--Butbaia--H\"ubsch--Jejjala--Mayorga Pe\~na--Mishra--Tan},colorlinks,citecolor=blue]{hyperref}
\usepackage{graphicx}
\usepackage{mathrsfs}
\usepackage{physics}
\usepackage{enumitem}
\usepackage[font={small,sf}]{caption, subcaption}
\usepackage[usenames,dvipsnames]{xcolor}
\usepackage{xfrac}
 
\definecolor{darkgreen}{RGB}{0, 204, 102}
\makeatletter
\def\@fnsymbol#1{\ensuremath{\ifcase#1\or *\or \dagger\or \ddagger\or \mathsection\or \mathparagraph\or \Diamond\or \star\else\@ctrerr\fi}}
\makeatother
\renewcommand\section[1]{${}$\\\noindent\textit{#1.}\enspace}

\renewcommand\tilde{\widetilde}

\def\rd{{\mathrm{d}}}

\begin{document}
\title{Precision String Phenomenology}

\author{Per Berglund}
\email{Per.Berglund@unh.edu}
\author{Giorgi Butbaia}
\email{Giorgi.Butbaia@unh.edu}
\affiliation{Department of Physics and Astronomy, University of New Hampshire, Durham, NH 03824, USA}
\author{Tristan H\"ubsch}
\email{thubsch@howard.edu}
\affiliation{Department of Physics and Astronomy, Howard University, Washington, DC 20059, USA}
\author{Vishnu Jejjala}
\email{v.jejjala@wits.ac.za}
\affiliation{Mandelstam Institute for Theoretical Physics, School of Physics, NITheCS, and CoE-MaSS, University of the Witwatersrand, Johannesburg, WITS 2050, South Africa}
\author{Dami\'an Mayorga Pe\~na}
\email{damian.mayorga.pena@tecnico.ulisboa.pt}
\affiliation{CAMSGD, Department of Mathematics, Instituto Superior T\'ecnico, Universidade de Lisboa, 1049-001 Lisboa, Portugal}
\author{Challenger Mishra}
\email{cm2099@cam.ac.uk}
\author{Justin Tan}
\email{jt796@cam.ac.uk}
\affiliation{Department of Computer Science \& Technology, University of Cambridge, Cambridge CB3 0FD, UK}

\begin{abstract}
Calabi--Yau compactifications of the $E_8\times E_8$ heterotic string provide a promising route to recovering the four-dimensional particle physics described by the Standard Model.
While the topology of the Calabi--Yau space determines the overall matter content in the low-energy effective field theory, further details of the compactification geometry are needed to calculate the normalized physical couplings and masses of elementary particles.
In this work, we present numerical computations of physical Yukawa couplings in a number of heterotic models in the standard embedding and demonstrate the existence of natural hierarchies, a coveted feature in string model building.
\end{abstract}

\keywords{string phenomenology, heterotic compactification, Calabi--Yau threefolds, Yukawa couplings}

\maketitle

\section{Introduction}\label{sec:introduction}
From the beginning, the $E_8\times E_8$ heterotic string has offered the potential for reproducing all observed physics below the string scale~\cite{Gross:1985fr}.
Indeed, compactification on a Calabi--Yau threefold, $X$, supplies a path to four-dimensional $\mathcal{N}{=}1$ supersymmetric quantum field theories with chiral matter~\cite{Candelas:1985en}.
Constraining an $SU(3)\subset E_8$ sub-bundle to equal the holomorphic tangent bundle, $T_X$, the simplest such models employ the ``standard embedding'', which determines the number of generations of chiral particles in the low-energy spectrum by topological features of the Calabi--Yau space: 
$N_\text{gen} \,{:=}\, \frac12 |\chi| \,{=}\, |h^{1,1}{-}h^{2,1}|$.
The effective gauge group upon compactification is $E_6$, and fields in the
$\mathbf{27}$ irreducible representation correspond to $(2,1)$-forms; similarly, $\overline{\mathbf{27}}$s correspond to $(1,1)$-forms.
To write the effective action, we must determine the physical, normalized $\mathbf{27}^3$ and $\overline{\mathbf{27}}{}^3$ Yukawa couplings~\cite{Strominger:1985ks,Strominger:1985it,CANDELAS1988458, rBeast}.

On a Calabi--Yau threefold, $(2,1)$-forms are tangent to the $h^{2,1}$-dimensional
complex structure moduli space, the geometry of which is captured by the Weil--Petersson metric~\cite{Candelas:1989ug, Candelas:1989qn, rBeast}.
This metric provides the physical normalization for kinetic terms of fields in the $\mathbf{27}$ representation and the sought after Yukawa couplings.
The Weil--Petersson metric can itself be determined in two equivalent ways: either from a Kodaira--Spencer map~\cite{keller2009numerical} or by calculation of period integrals~\cite{Morrison:1991cd,berglund1994periods}, which is straightforward when $h^{2,1}\,{\sim}\,1$.
The normalized $\mathbf{27}^3$ Yukawa couplings can also be computed directly by selecting appropriate harmonic representatives.
This requires knowledge of the Ricci-flat Calabi--Yau metric, which is certain to exist by Yau's theorem~\cite{Yau:1978cfy} and approximated using neural networks following~\cite{Ashmore:2019wzb,Anderson:2020hux, Douglas:2020hpv,Jejjala:2020wcc,Douglas:2021zdn,Larfors:2021pbb,Larfors:2022nep,Berglund:2022gvm,Dubey:2023dvu,Anderson:2023viv,Hendi:2024yin}.
These advances also facilitate scanning for high curvature regions within the Calabi--Yau space~\cite{Berglund:2022gvm}, which enable uplifting these models to a physically realistic de~Sitter spacetime~\cite{Bento:2021nbb}.

In this letter, we present efficient and numerically precise computations of the physical, normalized Yukawa couplings in various models and explicitly verify that all the methods for calculating these agree; see also~\cite{Butbaia:2024tje}.
Together with similar advances within ``line bundle'' constructions~\cite{Constantin:2024yxh}, this opens an avenue for \emph{completing} the effective action by computing also the $\mathbf{27}{\cdot}\overline{\mathbf{27}}{\cdot}\mathbf{1}$ and $\mathbf{1}^3$~\cite{rBeast} as well as higher order couplings~\cite{Gray:2024xun}.

\section{Normalized Yukawa couplings}
\label{sec:Yukawacouplings}
The matter spectrum of the four-dimensional effective field theory upon compactification is given by zero modes of the Dirac operator on $X$.
The left chiral superfields are in one-to-one correspondence with elements in $H^1(X;V)$, where $V$ is a holomorphic vector bundle with a given structure group $G\subset E_8$.
The corresponding unnormalized trilinear couplings enter the superpotential weighted by a factor
\begin{equation}\label{eq:unnorm}
\tilde{\kappa}_{ijk} = \int_X \Omega\wedge \tilde\Omega(a_i,a_j,a_k)\,,
\end{equation}
where $\{a_i\}_{i=1}^{h^1(V)}$ supplies a basis for $H^1(X;V)$ and $\tilde\Omega (a_i, a_j, a_k)$ denotes the appropriate contraction with the nowhere vanishing holomorphic $(3,0)$-form $\Omega$ yielding a singlet of the structure group of $V$.
This expression is quasi-topological, as it only depends on the cohomology classes of the respective matter fields.

The matter field kinetic terms arise from the K\"ahler potential.
These are not canonically normalized, but are instead weighted by a K\"ahler matter field metric, $G_{a\overline{b}}$.
In order to find the physical Yukawa couplings, we must transform to a basis of canonically normalized matter fields:
we first rotate to the eigenbasis of $G_{a \overline{b}}$, find the corresponding eigenvalues $\{\lambda_i\}_{i=1}^{h^1(V)}$, and compute
\begin{equation}
\kappa_{ijk} = \frac{\tilde{\kappa}_{ijk}}{\sqrt{\lambda_i \lambda_j \lambda_k}\, \mathcal{V}} \,,
\end{equation}
where $\mathcal{V}$ is the volume of $X$.

\section{Computing the K\"ahler metric $G_{a\overline{b}}$}\label{sec:WPmetric}  
The Dolbeault cohomology of the holomorphic vector bundle $V \rightarrow X$ carries a natural differential operator, $\overline{\partial}_V$.
When $V$ is endowed with a Hermitian structure, define the bundle Hodge-$\star$ operator as a $\mathbb{C}$-antilinear isomorphism:
\begin{gather}
    \overline{\star}_V: \Omega^{p,q}_X \otimes V \rightarrow \Omega^{n-p, n-q}_X \otimes V^*\,.
\end{gather}
We can then construct the codifferential and Laplacian operators on $V$ as:
\begin{gather}
    \overline{\partial}_V^{\dagger} := - \overline{\star}_V \circ \, \overline{\partial}_V \circ \overline{\star}_V \,, \quad \Delta_V := \overline{\partial}_V \overline{\partial}_V^{\dagger} + \overline{\partial}_V^{\dagger} \overline{\partial}_V \,.
\end{gather}
A bundle valued $(p,q)$-form $\eta \in \Omega^{p,q}_X \otimes V$ is said to be harmonic if $\Delta_V \eta = 0$.
The K\"ahler matter field metric $G_{a\overline{b}}$ in the low-energy effective action may be computed as the canonical inner product between harmonic representatives of the Dolbeault cohomology of $V$, $a,b \in H^1(X;V)$:
\begin{equation}\label{eq:norm} 
G_{a\overline{b}}\sim \int_X a \wedge \bar{\star}_V b\,.
\end{equation}
This expression depends both on the Ricci-flat Calabi--Yau metric on the base manifold $X$ and the Hodge-$\star$ on the vector bundle $V$, whose connection solves the Hermitian Yang--Mills equation~\cite{rBeast}.
In the standard embedding, where $V=T_X$, the metric on $X$ does double duty as the fiber metric, and the Zamolodchikov field metric coincides (up to a conformal factor~\cite{Dixon:1989fj}) with the Weil--Petersson 
metric over complex structure moduli space~\citep{tian:1987, todorov:1989, Candelas:1989qn, Candelas:1989ug, Dixon:1989fj, CANDELAS199121}:
\begin{equation}
\mathcal{G}_{a\bar{b}}
:=\int_X a\wedge \bar{\star}_g b \,.
\end{equation}

To calculate the Weil--Petersson metric 
recall that $X$ is covered by an atlas $\{ U_i \}$.
On overlaps $U_i \cap U_j$, coordinates are mapped by biholomorphic transition functions: $f_{ij}\,{:}\; z_j \mapsto z_i$.
To account for complex structure deformations and their effect on the transition maps, we write $f_{ij}(z_j;t)$ with $t=(t^1,\ldots,t^m)$, $m=h^{2,1}(X)$.
Locally, we interpret this construction as a fibration of $X$ over a base $B$ in the vicinity of a reference point $t_0$.
The Kodaira--Spencer map $\rho: T_{t_0} B \to H^1(X;T_X)$ is a map from the tangent space at $t_0$ to the first cohomology of $X$ with coefficients in $T_X$~\cite{56511be9-71c1-3ccb-89df-a73bcdcc07ed, kodaira_2005}.
Explicitly, the map is given by
\begin{gather}\label{eq:ks_map}
    \rho\left(\frac{\partial}{\partial t}\right) = \left[\left\{\frac{\partial f_{ij}^\mu(z_j, t)}{\partial t}\frac{\partial}{\partial z^\mu_{i}}\right\}\right],\quad\text{where}~z_j = f_{ji}(z_i;t) \,.
\end{gather}
Let $\mathcal{H}: H^p(X; T_X) \rightarrow H^p(X;T_X)$ denote the projection map sending representatives of $H^1(X; T_X)$ to their unique harmonic counterparts.
Then we obtain the following expression for the Weil--Petersson metric~\citep{tian:1987, todorov:1989}:
\begin{gather}
\mathcal{G}_{a\bar{b}} 
 = - \frac{\displaystyle \int_{X_t} \Omega(\mathcal{H}\rho(\partial / \partial t^a)) \wedge \overline{\Omega(\mathcal{H}\rho(\partial / \partial t^b))}}{\displaystyle \int_{X_t} \Omega \wedge \overline{\Omega}} \int_{X_t}\rd\mkern1mu\textsf{vol}_{g_t}\,.
 \label{eq:WPH}
\end{gather}
Here, $\Omega\left( \cdot \right)$ denotes the interior product with $\Omega$, which induces the isomorphism $H^1(X,T_X) \simeq H_{\overline\partial}^{2,1}(X)$.
Note that~\eqref{eq:WPH} requires knowledge of the Ricci-flat Calabi--Yau metric in order to compute the harmonic projection.
We now consider the effect of variations of complex structure on $\Omega_t$.
One can show that~\citep{keller2009numerical}:
\begin{gather}\label{eq:domegaDecomposition}
	\frac{\textrm{d}\Omega_t}{\textrm{d}t}\bigg\vert_{t_0} 
 = 	\Omega' + \Omega(\phi) ~\in~
 \Gamma(X, \Omega^{n,0}) \oplus \Gamma(X, \Omega^{n-1,1}) \,,
\end{gather}
where $\phi \in H^1(X;T_X)$ is a (not necessarily harmonic) representative of the Kodaira--Spencer class obtained via~\eqref{eq:ks_map}.
Similarly, $\Omega'$ is not holomorphic in general.
Both terms in this decomposition may be evaluated in terms of Poincar\'e residues~\citep{keller2009numerical, Butbaia:2024tje}.
Let $(-,-)$ denote the standard intersection pairing on $H^{p,q}_{\overline{\partial}}(X)$ with $p+q=n$:\begin{gather}
	(\alpha,\beta) = \int_X \alpha\wedge \overline{\beta}\,.\label{eq:blah}
\end{gather}
The Weil--Petersson metric on the complex structure moduli space can be directly obtained as
\begin{multline}\label{eq:WP_and_domega}
\mathcal{G}_{a\bar{b}}
         ~=~
  \left(\frac{\rd\Omega_t}{\rd  t^a}, \frac{\rd\Omega_t}{\rd t^b}\right)\bigg\vert_{t_0} \\
    + \frac{1}{(\Omega,\Omega)}\left(\Omega, \frac{\rd\Omega_t}{\rd t^a}\right)\bigg\vert_{t_0} \cdot  \overline{\left(\Omega, \frac{\rd\Omega_t}{\rd t^b}\right)}\bigg\vert_{t_0}\,.
\end{multline}
The problem is then reduced to the calculation of various integrals over $X$, numerically computed via Monte Carlo integration in local coordinates, as in~\citep{braun2008calabi}.
This provides a simple method for computing the field normalization, which does not require the harmonic projection and hence sidesteps explicit knowledge of the Ricci-flat metric on $X$.
This is a crucial technical simplification.

The Weil--Petersson metric can alternatively be constructed from explicit period integrals over $3$-cycles~\citep{CANDELAS199121,hosono1995mirror,Demirtas:2023als}.
As the periods satisfy a Picard--Fuchs equation~\cite{Morrison:1991cd}, calculating the Weil--Petersson metric in this fashion is practical only when there are a small number of complex structure moduli.

\section{Machine learning harmonic forms}
\label{sec:mlmetricsandharmonics}
The generic problem we consider is to approximate tensor fields on a space $X$ of nontrivial topology whose evolution is governed by a set of given partial differential equations (PDEs) on $X$.
We formulate this as a machine learning problem by:
    ({\small\textbf{1}}) reduction of the problem to learning a (possibly vector-valued) globally defined function $u$ over $X$;
    ({\small\textbf{2}}) expressing the condition for the relevant PDEs to be satisfied locally in terms of minimization of some functional $\mathscr{L}[u]$;
    ({\small\textbf{3}}) representing $u$ by a parameterized function, $u_{\theta}$ and subsequently minimizing the functional $\mathscr{L}[u_{\theta}]$ with respect to the parameters $\theta$ via gradient descent.

Numerical computations on manifolds are performed on local coordinate charts.
A tensor field $\mathscr{T}$ is a globally defined geometric object and hence must transform appropriately between coordinate patches.
In other words, the coordinate presentations $\mathscr{T}^{(i)}, \mathscr{T}^{(j)}$ in the patches $U_i, U_j$ over $X$ must be related by the transition functions appropriate to the tensor bundle.
The methods we present below ensure that the approximate tensor fields are globally defined by construction.
As we shall see, and has been previously observed~\citep{Berglund:2022gvm}, this is critical to ensuring the outputs of machine learning models are physically sensible.

The approach for learning the Calabi--Yau metric on $X$ is based on the spectral network construction~\cite{Berglund:2022gvm}.
It employs a preliminary transformation taking points in the manifold and turning this information into $\mathbb{C}^*$-invariant data, which is inputted to a feedforward neural network.
This ensures the output of the spectral network is well-defined on the ambient space $A$.
The final output is a scalar function $\phi_\text{NN}$, which by construction is a global function over the Calabi--Yau manifold.
The Ricci-flat metric approximation is~\citep{Larfors:2022nep}:
\begin{equation}
    g=g_s+\partial \bar\partial \phi_\text{NN} \,,
\end{equation}
with $g_s$ a reference K\"ahler metric in the desired class.
The optimal metric is the one that minimizes the loss: 
\begin{equation}
\ell_g = \int_X \rd\mkern1mu \textsf{vol}_\Omega\, \left|1-\frac{{\rm det}(g) }{\Omega \wedge \bar{\Omega}}\right| \,.
\end{equation}
We develop a similar approach to construct the harmonic bundle-valued $1$-form representatives.
We first use the Kodaira--Spencer map~\eqref{eq:ks_map} to compute $\Phi\in H^1(X;T_X)$; its harmonic projection may be approximated as
\begin{equation}\label{eq:harmonicNN}
    \mathcal{H}\Phi \approx \eta := \Phi+\overline{\partial}_{V} s_\text{NN} \,,
\end{equation}
with $s_\text{NN}$ a section of the holomorphic tangent bundle $T_X$~\cite{Ashmore:2021rlc,Douglas:2024pmn},
\begin{equation}\label{eq:sectionNN}
    s_\text{NN}=\sum_{ijkl} \psi_{ijkl}^{(\text{NN})}\frac{\overline{\alpha^{ij}_\nu}z^k z^l}{|z|^4}g^{\mu\bar{\nu}} \cdot \pdv{}{z^{\mu}} \in \Gamma(T_X)\,.
\end{equation}
Latin and Greek indices run over coordinates in $A$ and $X$, respectively.
Here, $\alpha^{ij}_\nu\, dz^\nu=\imath^*(z^i dz^j-z^i dz^j)$ is the pullback of basis of twisted $1$-forms on the ambient space $A$ along the inclusion map $\imath: X \hookrightarrow A$.
The coefficients $\psi_{ijkl}^{(\text{NN})}$ are produced as the output of a spectral neural network --- a vector-valued globally defined function over $X$.
This renders $\mathcal{H}\Phi$ a globally defined bundle-valued $1$-form by construction.
The K\"unneth formula permits a natural generalization of~\eqref{eq:sectionNN} when $A$ is a product of projective spaces.

Harmonic forms admit a natural variational interpretation as those forms which minimize the inner product $\left(\eta, \Delta_V \eta\right)$.
Our hypothesis~\eqref{eq:harmonicNN} is $\overline{\partial}_V$-closed by construction, and hence the natural variational objective to employ is simply the norm of the codifferential,
\begin{equation}\label{eq:harmonic_objective}
\ell_H = \left(\bar{\partial}_V^{\dagger} \eta, \bar{\partial}_V^{\dagger} \eta\right)\,.
\end{equation}
Notice that computation of the codifferential requires knowledge of the Ricci-flat metric on $X$, and that the harmonic forms must be computed via a two-stage process.
We first fix the complex structure moduli $t$ and approximate the Ricci-flat metric on $X_t$.
Secondly, we fix the learned metric on $X_t$, parameterize the $T_X$-section $s_{\text{NN}}$ with parameters $\theta$, and optimize the objective~\eqref{eq:harmonic_objective} with respect to $\theta$.

In summary, we have three distinct numerical methods of calculating the Weil--Petersson metric on moduli space and therefore the normalized Yukawa couplings.
The first uses the machinery of special geometry and works for any Calabi--Yau realizable as a complete intersection or toric variety; see~\eqref{eq:WP_and_domega}.
The second is based on a period integral construction, but is only practical for Calabi--Yau spaces with $h^{2,1} \sim 1$.
These first two methods are numerically exact (modulo integration error).
The third uses machine learning to approximate the bundle-valued harmonic forms corresponding to the matter fields of interest, and computes the Weil--Petersson metric via~\eqref{eq:WPH}.
We shall now illustrate these methods in various examples~\citep{supp}.
All three approaches exhibit excellent agreement, even close to singularities in the complex structure moduli space.

\section{Example~1: Mirror of $\mathbb{P}^5[3,3]$}\label{sec:Mirrorexample}
\begin{figure}[htb]
\centering
\includegraphics[width=1\linewidth]{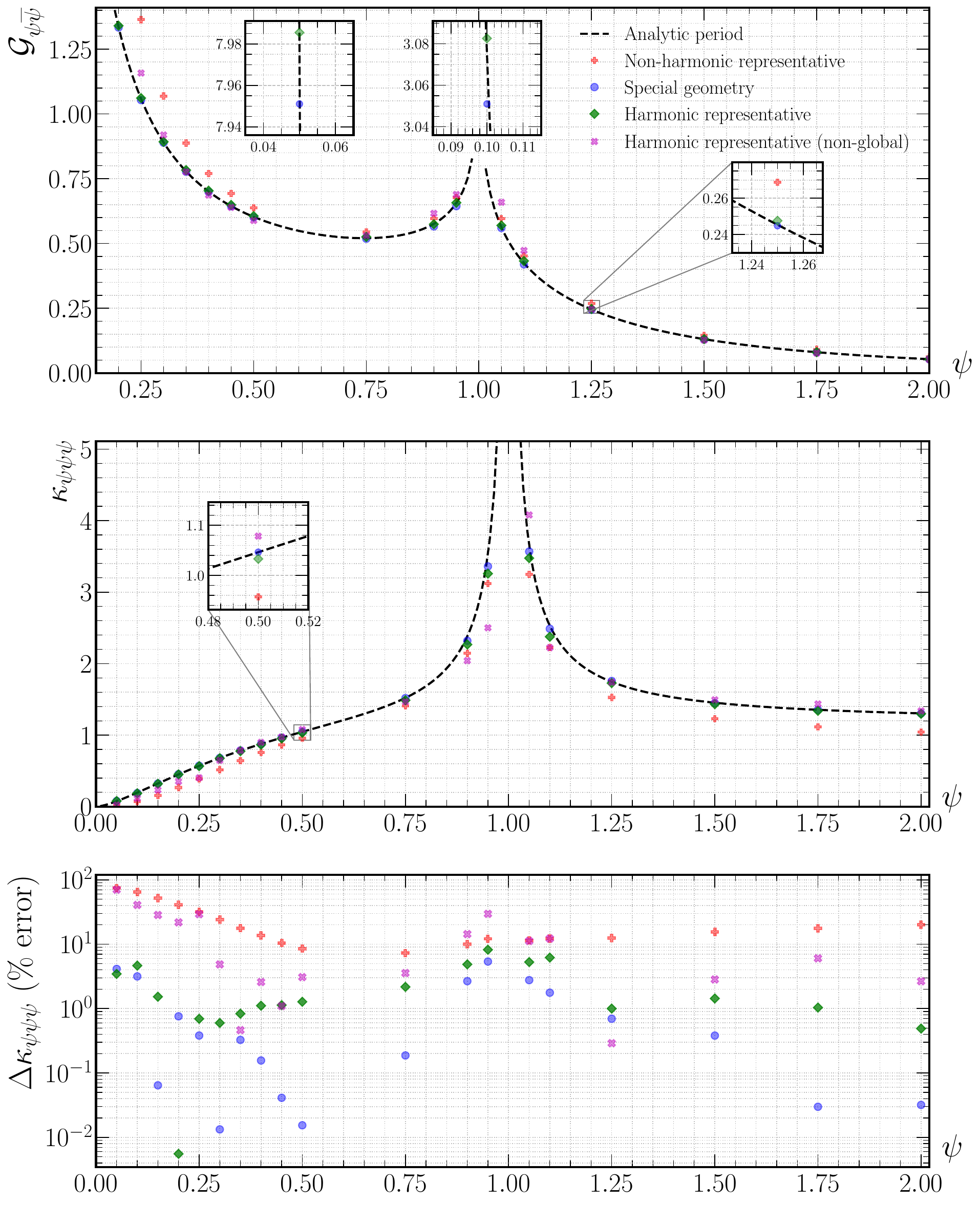}
\caption[X33comp]{Weil--Petersson metric (top), normalized Yukawa coupling (middle), and error in normalized Yukawa coupling compared to analytic expression from period integral (bottom) for the mirror of $\mathbb{P}^5[3,3]$ along the $\text{Im}(\psi)\,{=}\,0$ line in complex structure moduli space; the singular case of~\eqref{2cubics} at $\psi\,{=}\,0$ is at infinite distance in the moduli space.}
\label{fig:X33_comp}
\end{figure}

Consider the one-parameter family of complete intersections of two cubics:
\begin{equation}
\sum_{a=0}^2 x_a^3 - 3\psi x_3 x_4 x_5 
= \sum_{a=3}^5 x_a^3 - 3\psi x_0 x_1 x_2 = 0 \,, 
\label{2cubics}
\end{equation}
where $x_a$ are coordinates on $\mathbb{P}^5$ (\textit{cf.},~\cite{Libgober:1993aa, Joshi:2019nzi}) and $\psi$ deforms the complex structure.
Each member of this family is a Calabi--Yau threefold with $h^{1,1}=1$ and $h^{2,1} = 73$.
Their mirror is a blowup of a finite quotient of the same zero locus, and has $\widetilde{h^{1,1}}=73$ and $\widetilde{h^{2,1}}=1$.
Now, $\psi$ parametrizes the one non-blowup deformation of the complexified K{\"a}hler class.
We compute the Weil--Petersson moduli space metric $\mathcal{G}_{\psi\overline{\psi}}$ for the mirror family and the corresponding lone Yukawa coupling in Figure~\ref{fig:X33_comp}.

Here we evaluate the period computation on a fine grid along $\text{Im}(\psi)=0$, for $\psi \in [0,2]$.
We select a coarser grid over the same region and evaluate four different numerical methods for computing $\mathcal{G}_{\psi\overline{\psi}}$:
({\small\bf1})~Non-harmonic representatives (red) obtained from the Kodaira--Spencer representatives~\eqref{eq:ks_map}.
The values shown are the result of the computation in~\eqref{eq:WPH} without the harmonic projection.
({\small\bf2})~The special geometry computation (blue)~\eqref{eq:WP_and_domega}.
({\small\bf3})~The harmonic representatives (green) obtained via machine learning.
The values shown are the result of the computation in~\eqref{eq:WPH}, with the harmonic projection approximated via~\eqref{eq:harmonicNN}.
({\small\bf4})~The non-global harmonic representatives (red) are the output of a similar computation as the harmonic representatives, except $s_{\text{NN}}$ in~\eqref{eq:harmonicNN} is not inherently globally defined but is instead the direct output of a neural network.

The computation involving the globally defined approximate harmonic representatives is in excellent agreement with the special geometry computation and analytic period results, even close to both moduli space singularities, with errors of order $\sim 1\%$.
This demonstrates the importance of the hypothesis that $\eta$ in~\eqref{eq:harmonicNN} is globally defined by construction.
Notice also that as one approaches $\psi=0$, an infinite distance point in moduli space~\cite{Joshi:2019nzi}, the Yukawa coupling goes to zero.
This is another example supporting the observation of~\cite{Casas:2024ttx}, that in infinite distance limits certain families of Yukawa couplings vanish, as a signal of a tower of massive states becoming light.
For the case of the bicubic Calabi--Yau $\mathbb{P}^5[3,3]$, one can see that the point $\psi=0$ is a point of enhanced symmetry:
Considering the Abelian part only we observe a $\mathbb{Z}_3^4$ symmetry for any value of $\psi$.
This symmetry gets enhanced to $\mathbb{Z}_3^5\times \mathbb{Z}_6$ at $\psi=0$.
While no $\mathbb{Z}_3$ would forbid a trilinear coupling, it is the odd transformation of $\psi$ under the $\mathbb{Z}_2\subset \mathbb{Z}_6$ that enforces the vanishing of the coupling at the symmetric point.

\section{Example~2: A Tian--Yau quotient manifold}\label{sec:TianYauexample}
Consider the $1$-parameter family of Tian--Yau manifolds~\cite{Tian:1986ic} with the defining equations~\cite{Kalara:1987qv}:
\begin{equation}
\frac13 \sum_{a=0}^3 x_a^3 = \frac13 \sum_{a=0}^3 y_a^3 
= \sum_{a=0}^3 x_a y_a +\epsilon\sum_{a=2}^3 x_a y_a\,,
= 0 \,,
\label{e:TYe}
\end{equation}
where $x_a$ and $y_a$ are coordinates on $\mathbb{P}^3 \times \mathbb{P}^3$.
The Hodge numbers are $h^{1,1}=14$ and $h^{2,1}=23$.
Letting $\omega_3 = e^{2\pi i/3}$, there is a freely acting $\mathbb{Z}_3$-mapping:
\begin{equation}
\begin{array}{ccc}
(x_0,x_1,x_2,x_3) &\mapsto&
(x_0,\, \omega_3^{-1} x_1,\, \omega_3 x_2,\, \omega_3 x_3) \,, \cr
(y_0,y_1,y_2,y_3) &\mapsto& 
(y_0,\, \omega_3 y_1,\, \omega_3^{-1} y_2,\, \omega_3^{-1} y_3) \,.
\end{array}
\label{e:TYeZZ}
\end{equation}
For each $\epsilon$, the quotient by this discrete symmetry is a manifold with $\chi=-6$ and $h^{2,1} = 9$.
Thus, the standard embedding yields a three generation model~\cite{Tian:1986ic, Kalara:1987qv, CANDELAS1988357, candelas2008triadophilia} and supplies a useful testbed for phenomenological investigation.
A freely acting $\mathbb{Z}_3$ Wilson line does not suffice for the breaking of $E_6$ down to the Standard Model gauge group, $G_\text{SM} = SU(3)_C\times SU(2)_L\times U(1)_Y$.
Instead one has to produce an F- and D-flat configuration that Higgses the gauge symmetry to $G_\text{SM}$.
We refrain from this approach and instead consider the physically normalized Yukawa couplings of a trinification model.
The $23$ $\mathbf{27}$-plets split upon Wilson line breaking into the following massless representations under $SU(3)^3$~\cite{Greene:1987xh,Kalara:1987qv}: nine states denoted by $\lambda_i$ transforming as
$({\bf1},\overline{\bf3},{\bf3})$, seven states
$Q_i\sim ({\bf3},{\bf3},{\bf1})$, and seven states  $Q_i^c\sim(\overline{\bf3},{\bf1},\overline{\bf3})$.

It is possible to learn simultaneously all $h^{2,1}$ approximate harmonic forms representing the low-energy matter fields by extending~\eqref{eq:sectionNN} to have the spectral network produce a tuple of $h^{2,1}$ independent coefficients $\{\psi_{ijkl}^{(\alpha)}\}_{\alpha=1}^{h^{2,1}}$, with the variational objective now being the sum of the corresponding respective codifferential norms~\eqref{eq:harmonic_objective}.

\begin{figure*}[htb]
    \centering
    \includegraphics[width=1.0\textwidth]{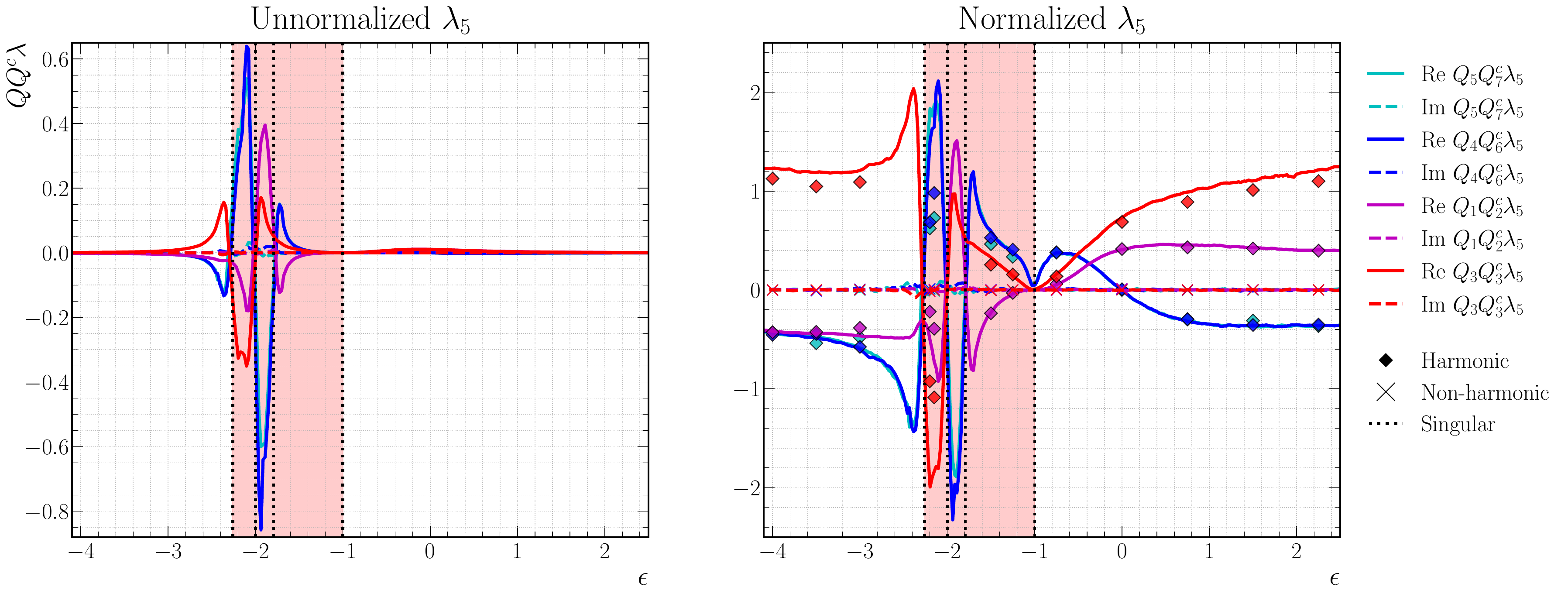}
    \caption{Unnormalized and normalized Yukawa couplings of form $QQ^c\lambda$ in the Tian--Yau quotient~\eqref{e:TYe}--\eqref{e:TYeZZ} showcasing the effect of physical normalization on the right, and exhibiting $\epsilon$-variable hierarchies near the conifold points along the $\Im(\epsilon) = 0$ slice in the shaded red region at $\epsilon \in \{{-}1$,
    ${-}\big(1{+}\sfrac1{\sqrt[3]2}\big)$,
    ${-}2$, 
    ${-}(1{+}\sqrt[3]2)$\}.}
    \label{f:QQl}
\end{figure*}

In Figure~\ref{f:QQl}, we illustrate the moduli space behaviour of select couplings for real values of the modulus $\epsilon$. The figure illustrates the effect of normalization: the physically normalized Yukawa couplings (right) markedly differ from their unnormalized values (left).
One can show that the Tian--Yau manifold becomes singular along the real $\epsilon$ line for $\epsilon \in \{-1,\,-1-2^{-1/3},-2,-1-2^{1/3}\}$ --- the region spanning those values is highlighted in red.
In particular we have considered couplings of the form $\lambda Q Q^c$ and have plotted nontrivial couplings involving the state $\lambda_5$.
Firstly, notice that the couplings $Q_5 Q_7^c \lambda_5$ and $Q_4 Q_6^c \lambda_5$ match for any value of $\epsilon$.
This equality is a consequence of an underlying $C$-symmetry exchanging them~\cite{Greene:1987xh}.
Due to the significant amount of discrete symmetries one has for this construction, it is actually observed that there are families of couplings that are equal (up to powers of $\omega_3$) as they are related by such discrete symmetries.
There is a change of sign in the coupling $Q_4 Q_6^c \lambda_5$ at $\epsilon=0$.
This is a finite distance point at which some hierarchies can be induced.
Furthermore, at $\epsilon=-1$, we observe that the Weil--Petersson metric becomes degenerate.
We notice that all couplings become zero, with a crossover of the $Q_1 Q_2^c\lambda_5$ coupling.
We numerically observe that the Weil--Petersson metric becomes divergent at the points in moduli space where the Tian--Yau manifold is singular, leading to substantial uncertainty around the normalized Yukawa couplings in this region arising from both Monte Carlo integration error and machine learning approximation error.

\begin{figure*}[htb]
    \centering
    \includegraphics[width=1.0\textwidth]{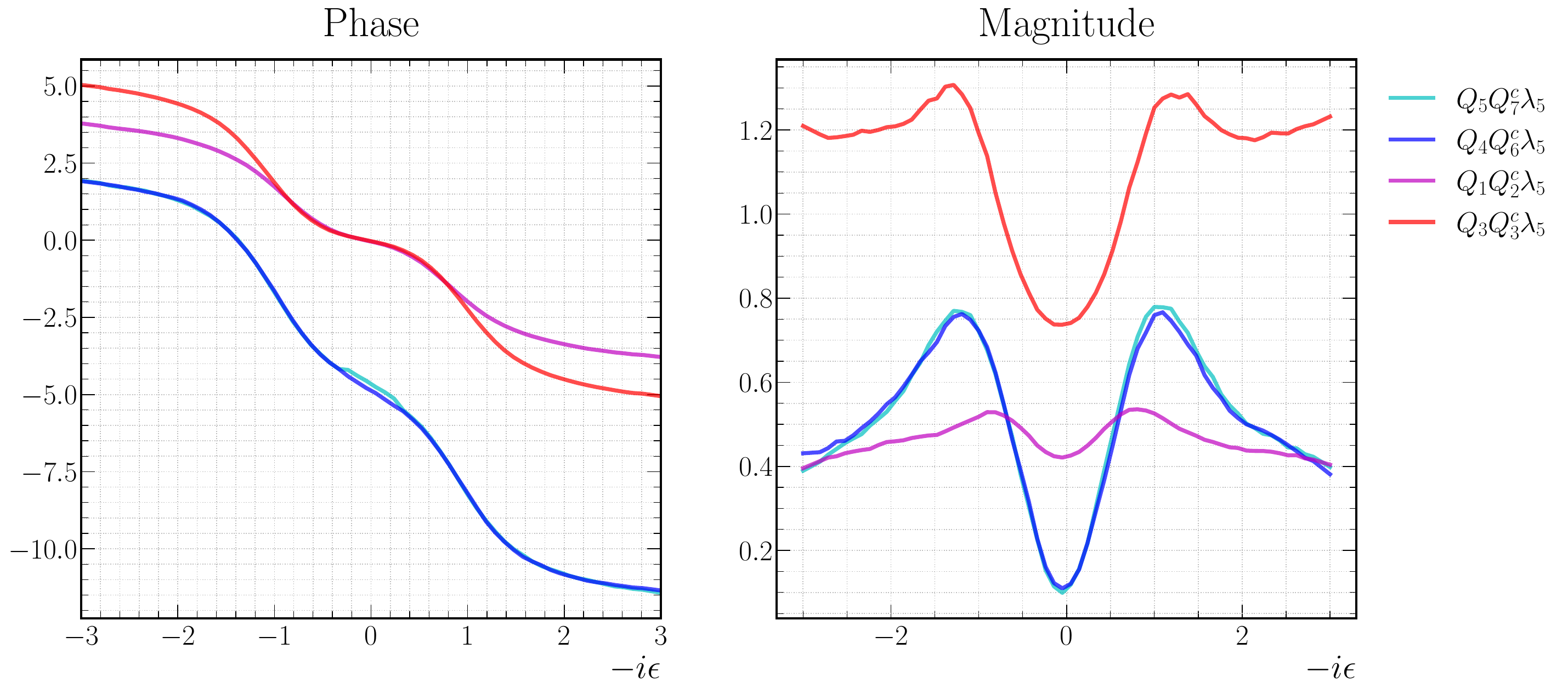}
    \caption{Normalized Yukawa couplings of form $QQ^c\lambda_5$ in the Tian--Yau quotient, including their relative phases (left) and magnitude (right) along purely imaginary axis, $\Re(\epsilon)=0$.
    $Q_5Q^c_7\lambda_5=Q_4Q^c_6\lambda_5$ continues to hold, maintaining the $C$-symmetry from~\cite{Greene:1987xh}.
    }
    \label{f:QQl-symmetrized}
\end{figure*}

Next, we turn to the possibility of physical CP violating phases arising in this scenario.
The eigenbasis may be modified by addition of phases to each physical state $\lambda_i\rightarrow e^{{\rm i}\phi_i}\lambda_i$ and similarly for $Q_i$ and $Q_i^c$.
This freedom can be used to reabsorb the phases in cubic Yukawa couplings of the form $\lambda_i^3$.
Once this procedure is performed, the physically relevant phases can be read off from the mixed couplings $\lambda_i \lambda_j \lambda_k$, $\lambda_i Q_j Q_k^c$ and so on.
Notably, all couplings are real along the real $\epsilon$-line.  Figure~\ref{f:QQl-symmetrized} exhibits the complex variation in the $\Re(\epsilon)=0$ direction. The phase plot on the right hand side of Figure~\ref{f:QQl-symmetrized} demonstrates that couplings have inequivalent phases and hence for $\Im(\epsilon)\neq 0$ there are opportunities to produce physically relevant CP violating phases.

\section{Discussion}\label{sec:discussion}
The machinery for the special geometric approach for computing the Weil--Petersson metric was developed in~\cite{keller2009numerical} for Calabi--Yau hypersurfaces.
In this work and in~\cite{Butbaia:2024tje}, we have generalized the theory to all complete intersection Calabi--Yau manifolds.
In particular, we are able to efficiently compute the normalized Yukawa couplings for arbitrary values of $h^{2,1}$.
Because the Weil--Petersson method only refers to the metric on the complex structure moduli space, the calculation is independent of the K\"ahler moduli.
Thus, the physical Yukawa couplings are the same in the standard embedding when computed from harmonic representatives using the Ricci-flat metric in any K\"ahler class.

We have further implemented a machine learned ansatz for the harmonic representatives that, in combination with the ``spectral network'' Calabi--Yau approximation, permits a separate computation of the Yukawa couplings.
The different methods employed yield the same numeric results and closely match the exact calculation where available.
This concordance proves the machine learning methods implemented~\cite{cymyc} to be robust, and suggests that this analysis may be generalized to arbitrary stable holomorphic vector bundles for which there is no known analog to special geometry.

In this letter, we have established that hierarchies in the Yukawa couplings are visible already with small excursions from symmetric points in complex structure moduli space.
Approaching phenomenologically suitable models entails providing Yukawa textures with a signature feature, namely, the presence of flavor hierarchies.
It is remarkable that the physically normalized Yukawa couplings allow for such hierarchies, which have been noticed for special points in moduli space.
In these situations, the couplings produced can be hierarchical at tree level.
This offers new possibilities beyond the Standard Model building paradigm where only large couplings are allowed at tree level while small couplings are perturbatively forbidden and effectively induced as powers of vacuum expectation values entering higher order couplings.

These advances pave the way for precise and efficient computations, and will play a role in further narrowing the search for phenomenologically appealing string compactifications.
Of course, the low-energy effective theory of any string compactification has a plethora of moduli fields which must be stabilized.
Since the full gauge group ($E_8$) is typically broken to the Standard Model gauge group via an intermediary GUT, considerations of symmetry breaking play an important role in describing the four-dimensional low-energy effective theory.
The standard embedding considered here leads to the gauge group $E_6$, the commutant of $SU(3)$ (\textit{viz.}, the holonomy of the Calabi--Yau threefold) in $E_8$.
Introducing Wilson lines makes other gauge groups possible.

Our results open up several intriguing avenues for future exploration.
Key questions about the string landscape, swampland conjectures, moduli stabilization, and the vacuum selection problem can be investigated more thoroughly using the tools developed in this work.

\section{Acknowledgements}
PB and GB are supported in part by the Department of Energy grant DE-SC0020220.
TH is grateful to the Department of Mathematics, University of Maryland, College Park, and the Physics Department of the Faculty of Natural Sciences of the University of Novi Sad, Serbia, for the recurring hospitality and resources.
VJ is supported by the South African Research Chairs Initiative of the Department of Science and Innovation and the National Research Foundation.
DM is supported by FCT/Portugal through CAMGSD, IST-ID, projects UIDB/04459/2020 and UIDP/04459/2020.
CM is supported by a fellowship with the Accelerate Science program at the Computer Laboratory, University of Cambridge.
JT is supported by a studentship with the Accelerate Science Program.
The authors would like to thank the Isaac Newton Institute for Mathematical Sciences for hospitality during the program ``Black holes: bridges between number theory and holographic quantum information'' (supported by EPSRC grant number EP/R014604/1) where work on this paper was partially undertaken.

\bibliographystyle{JHEP}
\bibliography{ref}

\end{document}